\documentstyle[12pt]{article}
\newcommand{\rf}[1]{(\ref{#1})}
\newcommand{\beq}{\begin{equation}}
\newcommand{\eeq}{\end{equation}}

\renewcommand{\o}{\omega}

\renewcommand{\S}{\Omega}
\newcommand{\bea}{\begin{eqnarray}}
\newcommand{\eea}{\end{eqnarray}}
\newcommand{\cD}{{\cal D}}
\newcommand{\cO}{{\cal O}}

\newcommand{\cC}{{\cal C}}

\newcommand{\noi}{\noindent}

\newcommand{\SL}{SL(2,{\bf Z})}

\renewcommand{\i}{^{-1}} 

\newcommand{\ver}{\thinspace\vert\thinspace}
\newcommand{\la}{\langle}
\newcommand{\ra}{\rangle}
\newcommand{\orbi}{\cC\wr \S}
\newcommand{\pri}{^\prime}
\newcommand{\CH}{\chi_p(x,y\ver\tau)}
\newcommand{\FR}{\displaystyle\frac}

\begin{document}
\topmargin 0pt
\oddsidemargin 5mm
\headheight 0pt
\topskip 0mm
\addtolength{\baselineskip}{0.20\baselineskip}
\pagestyle{empty}
\hfill 
\begin{center}
\vspace{3 truecm}
{\Large \bf Characters and modular properties of permutation orbifolds}
\vspace{5 truecm}

{\large Peter Bantay}
\vspace{1 truecm}

{\em Institute for Theoretical Physics\\
Rolland E\"otv\"os University, Budapest \\}

\vspace{3 truecm}
\end{center}
\noi
\underbar{\bf Abstract} 
Explicit formulae describing the genus one characters and
modular transformation properties of permutation orbifolds of
arbitrary Rational Conformal Field Theories are presented,
and their relation to the theory of covering surfaces is investigated.

\vspace{2 truecm}
\vfill
\newpage
\pagestyle{plain}

If $\cC$ denotes a Rational Conformal Field Theory, its $n$-th
tensor power $\cC^{\otimes n}$ is straightforward to describe for
any positive integer $n$, e.g. the primary fields are just $n$-tuples
of primaries of $\cC$, and their ( genus one ) characters are
simply the product of the corresponding $\cC$ characters. An interesting
feature of these theories is that any permutation $x\in S_n$ of
the $n$ "replicas" is a global symmetry of $\cC^{\otimes n}$, so
it is possible  to orbifoldize $\cC^{\otimes n}$ by any
permutation group $\S<S_n$.  For reasons to become
clear soon, we shall denote the resulting permutation orbifold by $\orbi$.

The first systematic investigation of permutation orbifolds has been 
performed in \cite{1} ( see also \cite{2} ). 
Permutation orbifold techniques have been applied in \cite{DMV2}
to compute the free energy of second quantized strings. 
Recently, a detailed
analysis of cyclic permutation orbifolds, i.e. the case $\S={\bf Z}_n$
for prime $n$, has been presented in \cite{BHS}, where the explicit
form of the genus one characters and their modular properties can
be found for $\S={\bf Z}_2$. The aim of the present paper is to generalize
the above results to arbitrary - possibly nonabelian - $\S$, and to
understand the underlying geometry. We shall only sketch the main results,
their derivation being left to a future publication.

The basic observation, which lies behind most of the results to be presented,
is that if  $\S_1\wr\S_2$ denotes the wreath product
of the permutation groups $\S_1$ and $\S_2$ ( c.f. \cite{ker}
), then the $\S_1\wr\S_2$ permutation orbifold of $\cC$ is 
nothing but the $\S_2$ permutation orbifold of $\cC\wr\S_1$, i.e.
\beq  \left(\cC\wr \S_1\right)\wr \S_2=
\cC\wr\left(\S_1\wr\S_2\right).\label{wr}\eeq
This property, which explains our choice for the notation,
is a straigthforward consequence of the definition of the
wreath product. In particular if $\cC$ is holomorphic,  i.e. it has only one
primary with respect to the maximally extended chiral algebra - e.g. 
the $E_8$ WZNW model at  level 1 -, 
the permutation orbifold $\cC\wr\S$ is a holomorphic orbifold model,
whose properties are described by the double $\cD(\S)$ of the group $\S$
\cite{DPR, B1}. 

A most important consequence of Eq. \rf{wr} is the
following description of the primary field content of permutation orbifolds :
the primary fields of $\orbi$ are in one-to-one correspondence with
the pairs $\langle p,\phi\rangle$, where $p$ is some
representative of an orbit of $\S$ acting on the $n$-tuples
$\langle p_1,\dots, p_n\rangle$ of primaries $p_i$ of $\cC$, 
while $\phi$ is an irreducible character of the
double $\cD(\S_p)$ of the stabilizer $$\S_p=\{x\in\S\ver xp=p\}$$ of
the $n$-tuple $p$.

If, as usual, $0$ denotes the vacuum of $\cC$, then the vacuum
of $\orbi$ - which we shall also denote by $0$ in the sequel -
corresponds to the pair $\la 0,\phi_0\ra$, where $[0]$ is the
one point orbit $\la 0,\dots ,0\ra$ whose stabilizer is
obviously $\S$ itself, while $\phi_0(x,y)=\delta_{x,1}$ is the
trivial character of $\cD(\S)$.

To write down explicitly the characters of the orbifold theory
$\orbi$, we have to introduce some notation. 
First, for a primary $p$ of $\cC$, we'll denote by 
$\chi_p(\tau)$ its genus one character, and by
$$\omega_p=\exp\left( 2\pi\imath(\Delta_p-\frac{c}{24})\right)$$
its exponentiated conformal weight, so that
$$\chi_p(\tau+1)=\omega_p\chi_p(\tau).$$
For a pair $x,y\in\S$ of commuting permutations, we shall
denote by $\cO(x,y)$ the set of orbits of the subgroup
generated by $x$ and $y$. To each ordered triple 
$\la x,y,\xi\ra$ with $\xi\in\cO(x,y)$, we
associate the following data :
\begin{enumerate}
\item $\lambda_\xi$ ( resp. $\lambda_\xi^*$ ) is 
the length of any  $x$ orbit ( resp. $y$ orbit ) contained in $\xi$ 
\item
$\mu_\xi$ ( resp. $\mu_\xi^*$ ) is the number of the $x$ orbits (resp. $y$
orbits ) 
\item $\kappa_\xi$ ( resp. $\kappa_\xi^*$ ) denotes the smallest non-negative integer 
for which $y^{\mu_\xi}=x^{\kappa_\xi}$ 
( resp. $x^{\mu_\xi^*}=y^{\kappa_\xi^*}$ ) holds
on the points of $\xi$. 
\end{enumerate}
These quantities are not independent, they are connected by the 
following relations : 

$$ \lambda_\xi \mu_\xi=\lambda_\xi^*\mu_\xi^*=\ver\xi\ver$$
\beq \mu_\xi^*=\gcd(\lambda_\xi,\kappa_\xi)\qquad {\rm and}\qquad
\mu_\xi=\gcd(\lambda_\xi^*,\kappa_\xi^*)\eeq
where $\ver\xi\ver$ is the length of the orbit $\xi$, and $\gcd(a,b)$ denotes
the greatest common divisor of the integers $a$ and $b$. 
\newpage

For an $n$-tuple $p=\la p_1,\dots p_n\ra$  of primaries of $\cC$
let's introduce the quantity

\beq \CH=\left\{\begin{array}{cl}{\displaystyle\prod_{\xi\in\cO(x,y)}\o_{p_\xi}^{-\kappa_\xi/\lambda_\xi}
\chi_{p_\xi}\left(\tau_\xi\right)}&\qquad\mbox{ if $x,y\in\S_p$ commute,}\\
0&\qquad\mbox{ otherwise}\end{array}\right.\label{CH}\eeq
where
\beq \tau_\xi=\frac{\mu_\xi\tau+\kappa_\xi}{\lambda_\xi}\label{tau}\eeq
and $p_\xi$ denotes the component of $p$ associated to the orbit $\xi$
( which is well-defined since $x,y\in \S_p$ ).
Then the character of the primary $\la p,\phi\ra$ of $\orbi$ reads
\beq \chi_{\la p,\phi\ra}(\tau)=\frac{1}{\ver\S_p\ver}
\sum_{x,y\in\S}\chi_p(x,y\ver\tau)\bar\phi(x,y)\eeq
Note that this formula is meaningful, i.e. it does not depend on the actual
representative $p$ of the orbit, since for all $z\in\S$
\beq \chi_{zp}\left(x,y\ver\tau\right)=\chi_p\left(x^z,y^z\ver\tau\right)\eeq 
In the special case $\S={\bf Z}_n$ we recover the results of \cite{BHS} for 
the characters of cyclic permutation orbifolds of prime order.

The geometry underlying the Eq.\rf{CH} is clear. If $E_\tau$
denotes a torus of modular parameter $\tau$, then
a commuting pair $x,y\in \S$ determines an $n$-sheeted unramified
covering of $E_\tau$, namely $x$ ( resp. $y$ ) describes how the sheets
are permuted when going around the $a$-cycle ( resp. $b$-cycle ) of a 
canonical homology basis. This covering is usually not connected, 
its connected components being in one-to-one correspondence with the orbits
$\xi\in\cO(x,y)$. By the Riemann-Hurwitz formula, each such connected
component is itself a torus $E_{\xi}$, 
whith modular parameter $\tau_\xi$  given by Eq. \rf{tau}.

In particular, if $Z(\tau)$ denotes the partition function of $\cC$, 
the partition function $Z_\S$ of $\orbi$ reads
\beq Z_\S(\tau)=\frac{1}{\ver\S\ver}\sum_{xy=yx}
\prod_{\xi\in\cO(x,y)}Z(\tau_\xi).\eeq

Once we know the explicit expression for the characters of $\orbi$,
we can compute their behavior under modular transformations.
Consider for example the $S$ transformation
\beq \tau\mapsto\tau\pri=-\frac{1}{\tau}\eeq
The modular parameter of the covering torus corresponding to the 
orbit $\xi\in\cO(x,y)$ will change accordingly 
\beq\tau_\xi\mapsto \tau_\xi\pri=
\frac{\mu_\xi\tau\pri+\kappa_\xi}{\lambda_\xi}=
\frac{\kappa_\xi\tau-\mu_\xi}{\lambda_\xi\tau}\label{A}\eeq
Besides acting on the modular parameter, $S$ also transforms the 
cycles in the canonical homology basis, thus changing the monodromy
of the covering. In the case at hand, this means that it transforms the
pair $\la x,y\ra$ into the pair $\la y,x^{-1}\ra$. This last pair
generates the same subgroup as $x$ and $y$, so that $\cO(x,y)=\cO(y,x\i)$,
but the covering torus corresponding to the triple $\la y,x\i,\xi\ra$ 
has modular parameter 
\beq\tilde\tau_\xi=\frac{\mu_\xi^* \tau- \kappa_\xi^*}{\lambda_\xi^*}\label{B}
\eeq
Clearly, $\tau_\xi\pri$ and $\tilde\tau_\xi$ should be related by a
modular transformation :
\beq \tau_\xi\pri=\frac{a_\xi\tilde\tau_\xi+b_\xi}{c_\xi\tilde\tau_\xi+d_\xi}
\label{C}\eeq
for some $S_\xi=\pmatrix{a_\xi&b_\xi\cr c_\xi&d_\xi\cr}\in\SL$, which 
- by  comparing Eqs. \rf{A} and \rf{B} - is easely seen to be
\beq S_\xi=\left(\begin{array}{ccc}\FR{\kappa_\xi}{\mu_\xi^*}&&  
\FR{\kappa_\xi\kappa_\xi^*-\mu_\xi\mu_\xi^*}{\ver\xi\ver}\cr \cr
\FR{\lambda_\xi}{\mu_\xi^*} && \FR{\kappa_\xi^*}{\mu_\xi}\end{array}\right)
\eeq

If we introduce the notation
\beq \Lambda_{pq}(M)=\o_p^{-a/c}M_{pq} \o_q^{-d/c}\eeq
for any $M=\pmatrix{a&b\cr c&d\cr}\in\SL$ with $c\ne 0$, then one checks that
\beq \chi_p\left(x,y \ver -\frac{1}{\tau}\right)=
\sum_q\prod_{\xi\in\cO(x,y)}\Lambda_{p_\xi q_\xi}\left(S_\xi\right)
\chi_q(y,x^{-1}\ver\tau)\eeq
which leads to the following formula for the matrix elements of the 
transformation $S$ in the orbifold theory $\orbi$ :

\beq 
S_{\la p,\phi\ra}^{\la q,\psi\ra}=\frac{1}{\vert\S_p\vert\vert\S_q\vert}
\sum_{z\in\S\atop x,y\in\S_p\cap\S_{zq}}\bar\phi(x,y)\bar\psi(y^z,x^z)
\prod_{\xi\in\cO(x,y)}\Lambda_{p_\xi}^{(zq)_\xi}(S_\xi)
\label{S}\eeq

As an example, consider the case $x=y=(1,2)$. Then an orbit
$\xi\in\cO(x,y)$ has either length $\ver\xi\ver=1$, in which 
case $S_\xi=\pmatrix{0&-1\cr 1&0}$ and $\Lambda(S_\xi)=S$,
or $\ver\xi\ver=2$, in which case $S_\xi=\pmatrix{1&0\cr 2&1}$
and 
\beq\Lambda(S_\xi)=T^{-1/2}S^{-1}T^{-2}ST^{-1/2}=
T^{1/2}ST^2ST^{1/2}\eeq
the latter equality being a consequence of the modular relation
$TSTST=S$. This is in complete agreement with the results of 
\cite{BHS} for the $S$-matrix of ${\bf Z}_2$ permutation orbifolds.

An important characteristic of a primary field is its $S$-matrix element with
the vacuum of the theory. From Eq. \rf{S} we get
\beq S_{0\la p,\phi\ra}=\frac{1}{\ver\S_p\ver}
\sum_{x\in\S_p}\phi(x,1)\prod_{\xi\in\cO(x,1)}S_{0p_\xi}\eeq
since for all $\xi\in\cO(x,1)$ we have $S_\xi=S$. 

Let's turn to the conformal weights, which can be determined from the behavior
of the characters under the modular transformation $T:\tau\mapsto\tau+1$.
We have
\beq \chi_p\left(x,y\ver\tau+1\right)=\chi_p(x,xy\ver\tau)
\prod_{\xi\in\cO(x,y)}\omega_{p_\xi}^{\mu_\xi/\lambda_\xi}\eeq
But it is straightforward to show that 
$\prod_{\xi\in\cO(x,y)}\omega_{p_\xi}^{\mu_\xi/\lambda_\xi}$ is independent
of $y$, so finally one gets
\beq \chi_{\la p,\phi\ra}(\tau+1)=\omega_{\la p,\phi\ra}\chi_{\la p,\phi\ra}
(\tau)\label{T}\eeq
where
\beq \omega_{\la p,\phi\ra}=\frac{1}{d_\phi}\sum_{x\in\S_p}\phi(x,x)
\prod_{\xi\in\cO(x,1)}\omega_{p_\xi}^{1/\vert\xi\vert}\label{weight}\eeq
is the exponentiated conformal weight
\beq \omega_{\la p,\phi\ra}=\exp\left(2\pi\imath(\Delta_{\la p,\phi\ra}-
\frac{nc}{24})\right)\eeq
of the primary ${\la p,\phi\ra}$ of $\orbi$,
and $d_\phi=\sum_{x\in\S_p}\phi(x,1)$.

This concludes our presentation of the structure of permutation orbifolds.
We have seen that the characters of these theories are completely determined
by the characters of the original theory and the action of the twist group.
The basic characteristics of the orbifold theory, such as the matrix elements
of the modular transformations $T$ and $S$ have been written down explicitly.
These latter determine through Verlinde's formula \cite{ver} 
the fusion rules of the permutation
orbifold. The connection with the theory of covering surfaces opens the way
to the computation of arbitrary correlation functions \cite{hv,dfms,adn}. 
Of course there's still
much to be done, e.g. one would like to have explicit expressions like Eqs.
\rf{S} and \rf{weight} for the fusion rules and the Frobenius-Schur indicators
\cite{FS}. 

{\it Acknowledgement :} It is a pleasure to acknowledge discussions with 
Zal\'an Horv\'ath, Geoff Mason, L\'aszl\'o Palla and Christoph  Schweigert.

\end{document}